\documentclass{aa}
\usepackage{epsfig}
\usepackage{natbib}
\newcommand{\kms}{\mbox{km s$^{-1}$}}

\newcommand{\hb}{\mbox{H$\beta$}}
\newcommand{\lya}{\mbox{Ly$\alpha$}}
\newcommand{\lam}{\mbox{$\lambda$}}

\newcommand{\forrefereestart}{\begin{bold}}
\newcommand{\forrefereeend}

\titlerunning{Compact Radio Sources and Jet-driven Feedback}
\authorrunning{Nesvadba et al.}

\begin{document}
\title{Compact radio sources and jet-driven AGN feedback in the early
  universe: Constraints from integral-field spectroscopy\thanks{Based on
    observations collected at the European Southern Observatory, Very Large
    Telescope Array, Cerro Paranal, Chile (076.A-0684(A))\newline
    $^\dag$Marie Curie Fellow,nicole.nesvadba@obspm.fr}}

\author{N.~P.~H. Nesvadba\inst{1,\dag}, M.~D.~Lehnert\inst{1}, C.~De
  Breuck\inst{2},  A.~Gilbert\inst{3}, W. van Breugel\inst{4}} 

\institute{GEPI, Observatoire de Paris, CNRS, Universite Denis Diderot, 5,
  Place Jules Janssen, 92190 Meudon, France 
\and
European Southern Observatory, Karl-Schwarzschild Strasse, D-85748 Garching
  bei M\"unchen, Germany
\and
Institute of Geophysics and Planetary Physics, Lawrence Livermore National
  Laboratory, 7000 East Avenue, L413, Livermore, CA 94550, USA
\and
University of California, Merced, P.O. Box 2039, Merced, CA 95344, USA
}

\date{Recieved / Accepted}

%\begin{abstract}
\abstract
%context
{}
%aims
{To investigate the impact of radio jets during the formation epoch of their
massive host galaxies, we present an analysis of two massive,
$\log{M_{stellar}/M_{\odot}} \sim 10.6$ and $11.3$, compact radio galaxies at
$z=3.5$, TNJ0205+2242 and TNJ0121+1320. Their small radio sizes (R$\le$ 10
kpc) are most likely a sign of youth. In particular, we compare their radio
properties and gas dynamics with those in well extended radio galaxies at high
redshift, which show strong evidence for powerful, jet-driven outflows of
significant gas masses (${M}\sim 10^{9-10}$ M$_{\odot}$).  
}
%method
{Our analysis combines rest-frame optical integral-field spectroscopy
obtained with SINFONI on the VLT with existing radio imaging, CO(4--3)
emission line spectra, and rest-frame UV longslit spectroscopy.}
%results
{[OIII]$\lambda$5007 line emission is compact in both galaxies and
lies within the region defined by the radio lobes.  For TNJ0205+2242,
the Ly$\alpha$ profile narrows significantly outside the jet radius,
indicating the presence of a quiescent halo. TNJ0121+1320 has two
components at a projected relative distance of $\sim$10 kpc and a velocity
offset of $\sim 300$ km s$^{-1}$, measured from the [OIII]$\lambda$5007
velocity map. This suggests that the fainter component is orbiting around
the more massive, radio-loud galaxy. If motions are gravitational, this
implies a dynamical mass of $2\times 10^{11} M_{\odot}$
for the radio-loud component.}
%conclusions
{The dynamical mass, molecular gas mass measured from the CO line
emission, and radio luminosity of these two compact radio galaxies imply
that compact radio sources may well develop large-scale, energetic
outflows as observed in extended radio galaxies, with the potential
of removing significant fractions of the ISM from the host galaxy. The
absence of luminous emission line gas extending beyond the radio emission
in these sources agrees with the observed timescales and outflow rates in
extended radio galaxies, and adds further evidence that the energetic,
large-scale outflows observed in extended radio sources (Nesvadba et
al. 2006) are indeed the result of influence of the radio jet.}

\keywords{cosmology: observations --- galaxies: evolution --- 
galaxies: kinematics and dynamics --- infrared: galaxies --- galaxies: jets}  

\maketitle 

\section{Introduction} 
\label{sec:introduction}
Although powerful AGN were the first sources discovered at high redshifts,
it is only now becoming clear that they may also have a profound impact
on the evolution of their host galaxies, particularly those at the upper
end of the galaxy mass function.  In order to explain some of the ensemble
properties of massive galaxies in the local Universe, such as the slope
of the upper end of the mass function \citep[e.g.,][]{benson03}, galaxy
evolution models have incorporated a phase of powerful AGN feedback in
the early evolution of massive galaxies \citep[e.g.,][]{silk98, hopkins05,
croton06, dimatteo05}. Such a phase effectively terminates star-formation
in already formed massive galaxies by heating and removing significant
fractions of the cold gas within the host galaxy.

However, direct observational evidence for strong AGN driven (negative)
feedback is still very rare. Although AGN undoubtedly maintain enormous
energy outputs over their lifetimes of $10^{7-8}$ yrs \citep{martini04},
we do not know by which physical mechanism this energy is transferred into
kinetic energy of the interstellar medium (ISM) of the host galaxy --
and whether any such mechanism exists that could provide the necessary
large coupling efficiencies. 

Radio jets in high-redshift radio galaxies are known
to strongly distort the gas kinematics along the jet axis
\citep[e.g.,][]{mccarthy96,tadhunter91,villar99}, and may therefore be
a candidate mechanism to ensure efficient coupling between the AGN
and the host galaxy -- if the influence of the jet is not confined
to the gas along the jet axis. Due to the complexity of the gas
dynamics in high-redshift radio galaxies, this is difficult to
determine from longslit spectra alone, but requires integral-field
spectroscopy. \citet{nesvadba06} find from rest-frame optical
integral-field spectroscopy of the z$=$2.2 powerful radio galaxy
MRC1138-262 that this may indeed be the case. They identify massive
outflows of $\sim 10^{9} M_{\odot}$ of ionized hydrogen, with relative
velocities of up to $>$ 2000 km s$^{-1}$ and emission line FWHMs $>
1000$ km s$^{-1}$, which do not appear to be strongly collimated along
the axis of the radio jet. However, \citeauthor{nesvadba06} find that
dynamical timescale and energy arguments nonetheless favor the radio
jet as being the dominant mechanism accelerating the gas, estimating
that over a lifetime of the outflow, t$_{dyn} \sim 10^7$ yrs, a total
of E$_{kin} \ga 10^{60}$ ergs of kinetic energy must have been injected
into the ISM to produce the observed kinematics. For powerful radio
galaxies like MRC1138-262, this energy is well above those typically
estimated for starburst-driven outflows, or even mergers, and
may be sufficient to unbind significant gas fractions from the dark
matter halo of a massive galaxy. Since z$\sim 2-3$ HzRGs appear
to be massive galaxies experiencing rapid growth of their stellar mass,
this suggests that jet-driven AGN feedback as observed in HzRGs may be a
promising mechanism to explain the particular characteristics of massive
galaxies at low redshift.

Compact high-redshift radio galaxies are particularly promising targets to
further investigate whether radio jets may cause large-scale outflows.
A radio source may be compact either as a result of a particularly
dense ISM, so that the radio jet takes a substantial amount of time to
break out, or perhaps not even break out of the ISM of the host galaxy
\citep[``frustrated jets'';][]{vanbreugel84}, or are of particularly
young ages \citep{blundell99}.  Radio spectral ages \citep[e.g.,
][]{murgia99} and dynamical timescales \citep{owsianik98} now indicate
that the latter appears to be the case; compact radio sources seem to
be particularly young, with ages of at most a few $\times 10^6$ yrs
\citep[see also][]{blundell99}. This makes them particularly interesting
for determining whether or not radio jets can drive large-scale outflows,
as the timescales of nuclear activity seem to be significantly shorter
than the timescales of starbursts and galaxy mergers \citep[$\sim 10^8$
yrs; e.g.,][]{barnes96}.

Compact radio galaxies may help us in several ways in deciding whether
radio jets are indeed the agents in generating the large-scale outflows
seen in extended powerful radio galaxies.  First, are the emission line
regions in these galaxies compact or extended? If the line emission in
these sources is related to a recent or on-going merger, then we do not
expect to find a correlation between the extended line emission and the
size and orientation of the radio source. In particular the emission
line region should not exclusively lie within the region defined by
the radio jets. Second, how do the kinematics of the emission line
gas relate to the region of the radio emission? Compact line emission
with similar kinematics as found in the extended radio galaxies and
lying within the radio emission will be a sign that the radio jets
play a major role for the gas kinematics in the compact source. If
the radio jet is the agent, then the kinematics of the emission line
gas outside of the region of radio emission should be quiescent, only
reflecting gravitationally driven motions. This is indeed what is being
observed in the rest-frame UV emission of extended radio galaxies at
high redshift with optical longslit and integral-field spectroscopy
\citep[e.g.,][]{villar03,villar07}. The reason is that the hotspot advance
speed is well beyond that expected for both a Compton heated bubble
or advance shocks, if either exist in these powerful AGN. Comparing
these galaxies to extended radio sources is crucial, as the absence of
high surface brightness gas, extended over several 10s of kpcs in the
compact sources will significantly strengthen the idea that the jets in
the extended radio galaxies entrained and accelerated significant gas
masses.  Consequently, we should observe smaller amounts of ionized gas
confined within the region of the radio emission in compact radio galaxies
because the timescales are so short in the compact sources that they have
not had time to entrain as much gas as the large scale radio sources.
Previous studies of compact radio sources based on longslit data
and at somewhat lower redshift, z$\sim 1$, suggested that this is indeed
the case \citep[e.g.][]{jarvis01,inskip02a,inskip02b,best00}.

Compelling evidence for radio jet-driven outflows in galaxies with compact
radio sources, sometimes with velocities of up to $V\ge 1000 $km s$^{-1}$
and multiple components with different kinematics, was found in longslit
and imaging spectroscopic studies at relatively low redshift \citep[z$\le
0.5$; e.g.,][]{tadhunter01,holt03,holt07,inskip07,tadhunter07}.
\citet{capetti99} used the Hubble Space Telescope Faint Object Camera
to obtain imaging and longslit spectra of the narrow-line region of the
low-redshift radio-loud Seyfert galaxy Mrk~3 at a spatial resolution of
25 pc.  They find strong evidence for radio jet-driven outflows in the
ionized gas. They propose that the gas is being accelerated and entrained
as the overpressurized cocoon of hot X-ray emitting gas surrounding
and energized by the radio jet expands.  However, at low redshift, the
accelerated gas masses, at least in ionized gas, do not appear sufficient
to have a large impact on the evolution of the host galaxy, as required in
cosmological models \citep{tadhunter07}, although a considerable fraction
of the total gas mass may be in other phases that are more difficult to
observe. \citet{morganti05} find outflows of neutral hydrogen implying
 mass outflow rates of up to 50 M$_{\odot}$ yr$^{-1}$, while, e.g.,
\citet{pounds03} argue for outflows of hot X-ray emitting gas.

Complementing these studies, and continuing our analysis of AGN--driven
outflows in radio galaxies at high redshift, we present an analysis of two
compact, steep-spectrum radio galaxies near the redshift where the co-moving
space density of powerful AGN was highest, z$\sim$ 2-4
\citep[e.g.,][]{willott01}. Direct high-redshift studies are essential in
understanding the interplay between the radio jets and the ISM at the epoch
when AGN were most powerful, massive galaxies were generally younger, and
where AGN likely had the greatest impact on galaxy evolution. With their
extreme star-formation rates of $\ge 1000$ M$_{\odot}$ yr$^{-1}$ and molecular
gas masses of a few$ \times 10^{10}$ M$_{\odot}$
\citep[e.g.,][]{papadopoulos00}, radio galaxies at z$\ge 1$ seem to have
fundamentally richer ISM than their lower-redshift counterparts, which appear
to be poorer in gas, particularly in molecular gas \citep{odea05}.

We obtained integral-field spectroscopy of the rest-frame optical emission
line gas of TNJ0121+1320 and TNJ0205+2242 at $z\sim 3.5$\footnote{Using
the flat $\Omega_{\Lambda}=0.7$ cosmology with H$_{0}=70$ km s$^{-1}$
Mpc$^{-1}$ leads to a luminosity distance D$_{L}=30.6$ Gpc and angular
size distance D$_{A}=$ 1.5 Gpc. The size scale is 7.3 kpc/\arcsec. The
age of the universe for this redshift and cosmological model is 1.8
Gyr.}, which were taken from the sample of \citet{debreuck00}. Both
are compact steep-spectrum sources with radii of 2 kpc and 10 kpc at
8.5 GHz, respectively. Their radio power at 1.4 GHz is very similar
$P_{1.4} = 2 \times 10^{25}$ W. K-band imaging of TNJ0121+1320 shows
two knots separated by $1.5$\arcsec, whereas TNJ0205+2242 appears
compact in the K-band, and is marginally extended along the axis of the
radio jet \citep{debreuck01b}. \citet{debreuck03} obtained CO(4--3)
emission line data on TNJ0121+1320. The molecular line emission does
not appear spatially resolved; the spatial resolution suggests a size
of $\le 4 $ kpc, while they find a line width of FWHM $\sim 700$ km
s$^{-1}$. \citet{debreuck01,humphrey07} discuss the Ly$\alpha$ emission
from the two sources. Both galaxies have luminous rest-frame UV emission;
TNJ0121+1320 appears compact, whereas TNJ0205+2242 appears surrounded
by a halo of narrow Ly$\alpha$ emission extending to $\ge 60$ kpc.

\section{Observations and data reduction} 
We observed TNJ0205+2242 and
TNJ0121+1320 in the K band with the integral-field spectrograph SINFONI
\citep{bonnet04} on the VLT in December 2005 under good conditions. Both
targets are at redshifts $z\sim 3.5$, so that our observations cover the
wavelengths of [OIII]$\lambda \lambda$4959,5007 and \hb. SINFONI is an
image-slicing integral-field spectrograph, with 8\arcsec$\times$ 8\arcsec\
field of view at a sampling of 0.25\arcsec$\times$ 0.25\arcsec\ pixel
scale, and spectral resolution of $R \sim 4000$ in the K band. Individual
exposure times were 600 s, and we obtained a total of 10800 seconds and
7200 seconds of on-source exposure time for TNJ0205+2242 and TNJ0121+1320,
respectively.

We used the IRAF \citep{tody93} standard tools for the reduction
of longslit-spectra, modified to meet the special requirements of
integral-field spectroscopy, and complemented by a dedicated set of IDL
routines. Data are dark-frame subtracted and flat-fielded. The position of
each slitlet is measured from a set of standard SINFONI calibration data,
measuring the position of an artificial point source. Rectification
along the spectral axis and wavelength calibration are done before
night sky subtraction to account for some spectral flexure between the
frames. Curvature is measured and removed using exposures of an arc lamp,
before shifting the spectra to an absolute (vacuum) wavelength scale
with reference to the OH lines in the data. To account for variations
in the night sky emission, we normalize the sky frame to the average of
the object frame separately for each wavelength before sky subtraction,
masking bright foreground objects, and correcting for residuals of the
background subtraction and uncertainties in the flux calibration by
subsequently subtracting the (empty sky) background separately from each
wavelength plane.

The three-dimensional data are then reconstructed and spatially aligned
using the telescope offsets as recorded in the header within the same
sequence of 5-6 dithered exposures (about one hour of exposure), and by
cross-correlating the line images from the combined data in each sequence
of exposures, to eliminate relative offsets between different sequences.
Telluric correction is applied to each individual cube before the
cubes are combined from the several sets of sequences taken for each
source. The relative flux calibrations are obtained from observations
of standard stars. From the light profile of the standard star (taken
after each sequence of exposures or one per approximately every hour),
we measure the FWHM spatial resolution in the combined cube.

TNJ0205+2242 is sufficiently close to an R=14.8 mag bright star, allowing
for adaptive optics-corrected observations. With strongly varying
coherence times, we obtained Strehl ratios $\sim 0.2-0.7$ (as given in
the frame headers), resulting in an overall spatial resolution of FWHM
= 0.45\arcsec$\times$0.4\arcsec\ in right ascension and declination in
the final cube, respectively. During our adaptive-optics assisted
observations with SINFONI, we monitored the point spread function by
making intermittant observations of a nearby bright star.  We used the
combined image of this star to estimate the size of the seeing disk
in our data. Our seeing limited data set of TNJ0121+1320 has a FWHM
resolution of 0.65\arcsec$\times$0.55\arcsec in right ascension and
declination respectively, measured from the PSF of the standard star taken
during the observations.

\subsection{Complementary data sets and relative astrometry} 
\label{ssec:alignment}
We complement our SINFONI data with the 8.4 GHz radio data of
\citet{debreuck02} for TNJ0121+1320, and new, deep 8.4 GHz VLA A-array maps of
TNJ0205+2242 with a spatial resolution of $\sim 0.4$\arcsec. We also use
rest-frame UV spectroscopy, which was initially described in
\citet{debreuck01}, to compare the rest-frame optical emission line
properties with those of the Ly$\alpha$ emission.

Studying the interplay between radio jets and the ISM of high-redshift
galaxies requires an alignment that is accurate to $\le$ 0.5\arcsec,
which is beyond the pointing accuracy of the VLT. TNJ0121+1320 is not
resolved at 8.4 GHz, and consists of 2 compact components in rest-frame
optical continuum and emission line images. Since the position of
the radio peak, taken at face value, falls close to the center of the
brighter peak in the K band image of TNJ0121+1320, we place the radio
source into the center of this component. Moreover, the redshift and
line width of the [OIII]$\lambda$5007 line emission of component $A$ are
the same as the CO(4--3) emission line redshift and line width measured
by \citet{debreuck03}, which also suggests that component $A$ dominates
the overall mass budget of the system and is more likely to host the AGN.

To align the radio and infrared data of TNJ0205+2242, we use the radio
core, which is detected in our deep A-array imaging at 8.4 GHz. In the
infrared data, we find an unresolved narrow line region, and broad
H$\beta$ emission at the same position. Since these are the typical
rest-frame optical features of an AGN, we place the radio core at the
peak of the narrow-line region.  Although this placement is justifiable
on astrophysical grounds and in analogy with what is known about radio
galaxies generally and other types of powerful AGN, we note that the exact
alignment of the radio at the level of a few tenths of an arcsecond has
no impact on the overall conclusions presented here.

\section{TNJ0205+2242}
\label{sec:tnj0205}

\subsection{[OIII] Emission line and continuum morphology} 
\label{sec:tnj0205.morph}
\begin{figure}
\centering
\epsfig{figure=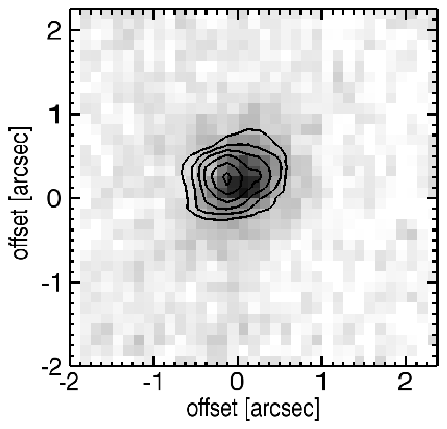,width=0.23\textwidth}
\epsfig{figure=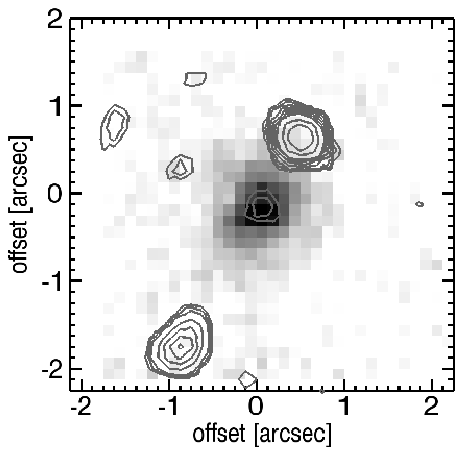,width=0.23\textwidth}
\caption{The left panel shows the [OIII]\lam5007 emission line morphology of
TNJ0205+2242 as grey scale with the continuum morphology overlaid as
contours. Contours are given in steps of 2$\sigma$, starting at 3$\sigma$. The
$\sim 0.25$\arcsec\ relative offset between the two continuum peaks is
significant, because both images were extracted from the same data cube. The
right panel shows the same [OIII]\lam 5007 emission line image, but the
contours now indicate the 8.4 GHz radio morphology. North is up, east to the
left in both images.}
\label{fig:0205images}
\end{figure}

We show the morphology of the [OIII]\lam5007 emission line regions of
TNJ0205+2242 in the left panel of Fig.~\ref{fig:0205images} with the
continuum morphology shown as contours. The line emission is spatially
offset from the continuum peak by about 0.25\arcsec\ to the south-west,
which is significant because both images were extracted from the same
data cube. Continuum emission is faint, K$ =21.6\pm$ 0.3 mag within a box
aperture with a length of 1.5\arcsec\ on either side. We use the
FWHM of the continuum image to approximate the radius of the continuum
emitting region and assume that the observed full width at half maximum
of the continuum is a quadratic sum of the intrinsic FWHM of the source
and the FWHM of the seeing disk.  We find a seeing corrected full width at
half maximum of the source of 0.6\arcsec, corresponding to about 4.5 kpc.

[OIII]\lam5007 line emission is only marginally resolved, and elongated
along an axis stretching from north-west to south-east, that is, along
the axis of the radio jet. Assuming that the intrinsic source FWHM and
the seeing FWHM add in quadrature, we estimate the FWHM of the emission
line region to be 1.06\arcsec $\times$ 0.7\arcsec\ (8 kpc $\times$5 kpc)
along the radio jet axis and perpendicular to it, respectively.

\subsection{Integrated spectrum} 
\begin{figure}
\epsfig{figure=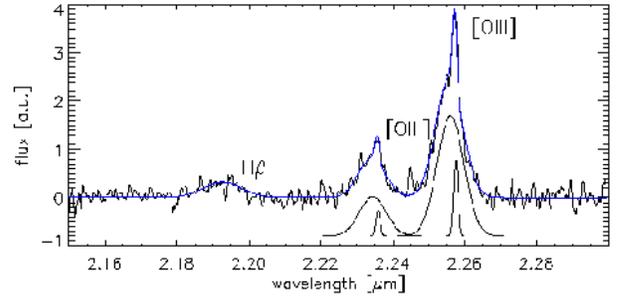,width=0.23\textwidth,angle=90}
\caption{Integrated K-band spectrum of TNJ0205+2242. The blue line
indicates the Gaussian fit to the line profile. Black Gaussian curves
(shifted by an arbitrary amount along the ordinate) show the
individual fit components.}
\label{fig:0205intspec}
\end{figure}

The integrated spectrum of TNJ0205+2242 is shown in
Fig.~\ref{fig:0205intspec}. The blue lines indicate Gaussian fits to the
line profiles. We constrained both [OIII]\lam4959,5007 lines to have the
same redshift and line widths, and imposed a flux ratio R$_{4959,5007}
= 0.33$. The [OIII]\lam4959,5007 doublet is well described with two
Gaussian components. The narrow component is at a redshift $z=3.50867 \pm
0.0007$, and has a FWHM$ = 143\pm 10$ km s$^{-1}$. The broad component
is blueshifted by $195\pm 5$ \kms\ relative to the narrow component,
with a width of FWHM$=1171\pm 71$ \kms.

\hb\ line emission is significantly broader than [OIII], although the
exact line properties are difficult to determine, due to the faintness of
the signal in each spectral bin, and several superimposed OH lines. 
The uncertainty is estimated from Monte-Carlo simulations taking into
account the signal to noise ratio, line width, and spectral binning, but not
the superposition of the OH line residuals, which are difficult to model
precisely. We therefore caution that we may somewhat underestimate the 
uncertainties of the H$\beta$ measurement, in particular the line width.

Our best estimate, based only on pixels that are not affected by OH lines,
implies a Gaussian width of FWHM$=1699 \pm 145$ \kms\ and a redshift
$z=3.50882 \pm 0.00086$, which is similar to the redshift of the narrow
[OIII]\lam5007 component within the uncertainties. 
The FWHM$\sim 1700$ km s$^{-1}$ may appear large for a high-redshift
galaxy, and is near the dividing line to quasars \citep[e.g.,][find that
nuclear broad lines in radio-loud quasars can be as narrow as FWHM$\sim 2000$
km s$^{-1}$ ]{sulentic00}. However, similar FWHMs have also been observed in
HzRGs \citep[][find an average FWHM(H$\beta$)$=$1530 km s$^{-1}$ in their
sample of 15 z$=2-2.6$ HzRGs]{iwamuro03}. This and the faintness of the radio
core in TNJ0205+2242 suggest that this galaxy is indeed a radio galaxy rather
than a quasar.

The line flux is $3.8 \pm 0.4 \times 10^{-17}$ erg s$^{-1}$ cm$^{-2}$. Broad
\hb\ line emission is detected within an area of approximately one seeing disk
surrounding the peak of the [OIII]\lam5007 emission line region shown in
Fig.~\ref{fig:0205images}. All line widths are corrected for the instrumental
resolution, which we determined from the measured width of night-sky lines
from a cube combined in the same way as for the cubes for the radio galaxy.

\subsection{Spatially resolved gas kinematics}  
\label{ssec:tnj0252extended}
\begin{figure}
\epsfig{figure=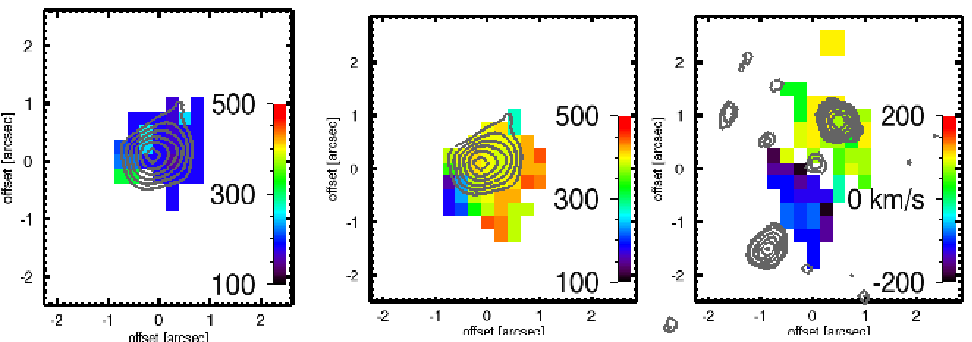,width=0.48\textwidth}  \\
\epsfig{figure=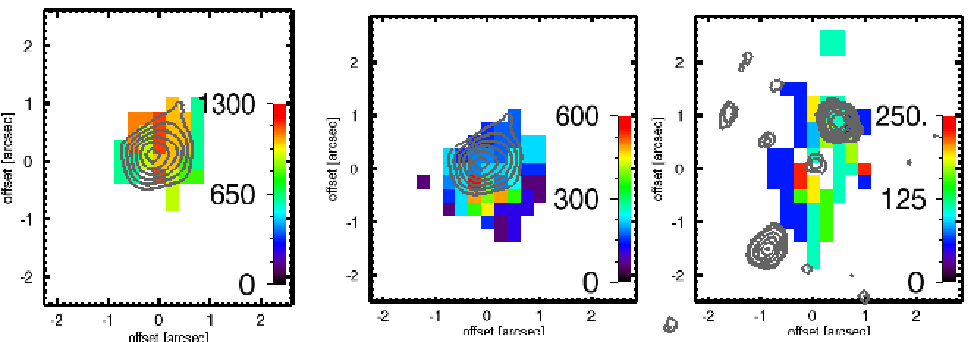,width=0.48\textwidth}\\
\epsfig{figure=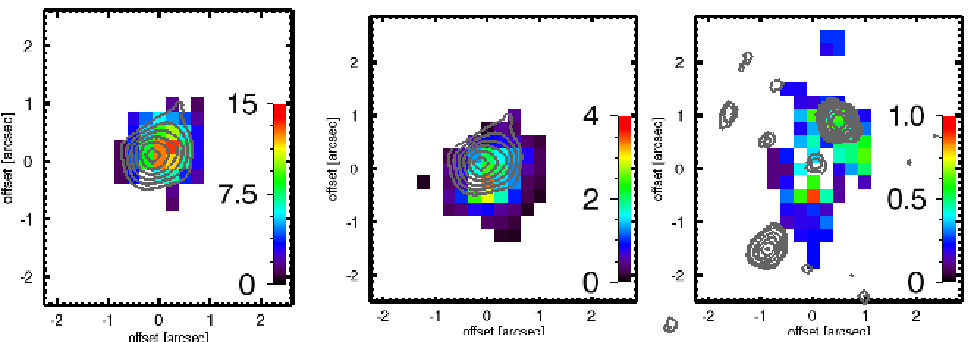,width=0.48\textwidth}
\caption{Maps of relative velocities (top), FHWMs (middle), and line fluxes
(bottom) of the 3 line components in TNJ0205+2242, all measured from
[OIII]$\lambda$5007. Contours indicate the K band continuum emission in the
left and the mid panel, respectively. Contours in the right panel indicate the
8.4 GHz emission. Colorbars show the velocities and FWHMs in km s$^{-1}$, and
fluxes (in arbitrary units), in the top, middle, and bottom panels, respectively. }
\label{fig:0205kinematics}
\end{figure}

The very regular integrated spectrum does not reflect an
overall uniform kinematics of emission line gas, as seen from
Fig.~\ref{fig:0205kinematics}. To construct these maps, we extracted
individual spectra from 5 pixels $\times$ 5 pixels (0.6\arcsec$\times$
0.6\arcsec) box apertures and simultaneously fitted line profiles with
up to 4 Gaussian components to the [OIII]\lam5007 emission line in each
spectrum, requiring S/N $>$ 5 for each component.

We identify three emission line regions with different kinematics and
morphologies. (1) A component with a biconal structure, which has a clear
($197\pm 47$ \kms) velocity difference between the two sides. It is
elongated along the axis of the radio jet and extends to the northern
radio lobe. Lines are narrow, with $\sigma \sim 118$ \kms. We will in
the following use the midpoint between the average velocity in the two
sides as reference velocity for the other components.  (2) The broadest
component is distributed symmetrically around the peak of the line image
in Fig.~\ref{fig:0205images} with a spatial extent of approximately one
seeing disk. The width of this component is FWHM$= 1083\pm 106$ \kms. The
uncertainty in this and the following estimates includes the 1 $\sigma$
scatter about the mean. Velocities are uniform in this component, v$ =
197\pm 37$ \kms.  (3) A narrow line component with a similar radius and
a range v$ = 389\pm 91$ \kms. Both regions (2) and (3) are spatially
unresolved and likely represent the narrow line region of the AGN.

\subsection{Constraining the mass of TNJ0205+2242} 
\label{ssec:tnj0205mass} 
Due to the complexity of the dynamics of high-redshift radio galaxies,
it is in most cases not possible to use their internal kinematics
for dynamical mass estimates, as those are not dominated by the
large-scale gravitational motion (but see \S\ref{ssec:tnj0121mass} for an
exception). However empirically, lines originating from the narrow line
regions in AGN are found to correlate with the dynamics of the bulge,
e.g. the [OIII]$\lambda$5007 luminosities \citep[e.g.,][]{kauffmann03} or
[OIII]$\lambda$5007 dispersions \citep[e.g.][]{nelson96}. Since absolute
flux calibrations in the near infrared are inherently uncertain, and we
are not able to account for extinction, we will in the following use the
[OIII]$\lambda$5007 emission line dispersion measured from the narrow
component of the integrated spectrum of TNJ0205+2242, $\sigma_{[OIII]}
= 62$ km s$^{-1}$, and the relationship of \citet{nelson96}. For the
measured [OIII] dispersion the \citeauthor{nelson96} correlation suggests
a stellar velocity dispersion of the host, $\sigma_{host}^{0205} \sim 85$
km s$^{-1}$, with a scatter between $\sim 55-136$ km s$^{-1}$. Two narrow
components contribute to the narrow line in the integrated spectrum
of TNJ0205+2242 (\S\ref{ssec:tnj0252extended}), but the total narrow
line flux is dominated by the spatially unresolved narrow-line region
(see Fig.~\ref{fig:0205kinematics}). Using the best-fit estimate of
\citet{nelson96} to approximate the velocity dispersion of TNJ0205+2242,
we give a rough mass estimate by setting,

\begin{equation}
\label{eqn:dispmass}  
M_{host}^{0205} = \frac{5\ \sigma^2\ R}{G} = 1.2\times 10^{10}\ \biggl(\frac{V}{100
km s^{-1}}\biggr)^2\ \frac{R}{kpc} [M_{\odot}] 
\end{equation} 
with the velocity dispersion $\sigma$, and radius $R$ of the host (which
we approximate by the half-light radius of the emission line region
of TNJ0205+2242, $R\sim 3$ kpc), and the gravitational constant, $G$.
For TNJ0205+2242, 
adopting the best-fit $\sigma$ in the \citeauthor{nelson96} correlation 
as given,
we estimate $M_{host}^{0205} \sim 3\times 10^{10}$
M$_{\odot}$. 
However, we caution that the astrophysical uncertainties of this
estimate are large including, e.g., the size estimate, scatter in the
\citeauthor{nelson96} correlation, and possible evolutionary
effects. Nonetheless, this suggests that a mass estimate in the range of few
$\times 10^{10} M_{\odot}$ is a reasonable estimate.

Photometric masses of HzRGs are also difficult to estimate, as the AGN
contaminates the continuum emission by a degree that is hard to constrain,
especially with ground-based observations. \citet{seymour06} obtained stellar
mass estimates for a sample of 69 HzRGs with compact and extended radio
sources, for which they obtained Spitzer IRAC and MIPS data, allowing them to
constrain the AGN contribution to the SEDs, and in particular through
measuring the 1.6$\mu$m peak of the stellar contribution, which is suitable to
the stellar mass of passively evolving systems with ages $> 1$ Gyr.  While
this may overestimate the age of HzRGs, it does provide an upper limit to the
stellar mass in these systems, since the mass-to-luminosity ratio will
increase as the stellar population grows older. For TNJ0205+2242 specifically,
they estimate $M_{stellar}^{0205} < 6\times 10^{10}$ M$_{\odot}$, which, given
the large uncertainties of either estimate is in good agreement with our
results. 

\section{TNJ0121+1320}
\subsection{[OIII] Emission line and continuum morphology} 
\label{ssec:0121morph}
\begin{figure}  
\epsfig{figure=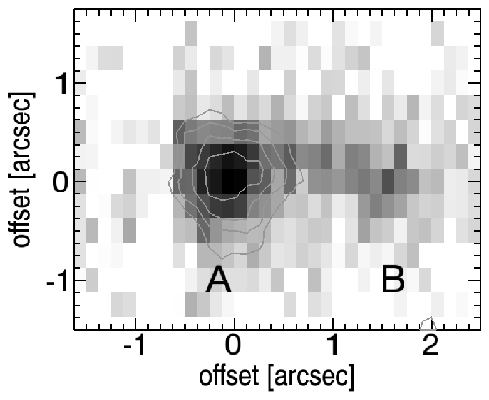,width=0.23\textwidth}
\epsfig{figure=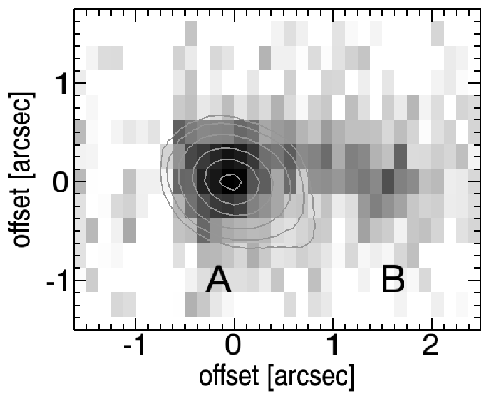,width=0.23\textwidth} 
\caption{The left panel shows the [OIII]\lam5007 emission line morphology of
TNJ0121+1320 as grey scale with the continuum morphology overlaid as
contours. Contours are shown in steps of 2$\sigma$, starting from
3$\sigma$. The right panel shows the same [OIII]\lam 5007 emission line image,
but the contours now indicate the 8.4 GHz radio morphology. North is up, east
to the left in both images.
}
\label{fig:0121images}
\end{figure}

The morphology of the [OIII]\lam5007 emission line region of the second
source, TNJ0121+1320, is shown in Fig.~\ref{fig:0121images}. The
line-free continuum is shown as contours. TNJ0121+1320 consists of two
components in the line image. We will in the following refer to the
brighter eastern emission line region as component ``A'', and to the
fainter western region as component ``B''. 

The projected distance between the two components is $\sim 1.4$\arcsec,
corresponding to $\sim 10$ kpc. This is well above the seeing disk, which has
a FWHM$=$0.65\arcsec$\times$0.55\arcsec\ in right ascension and declination,
respectively. Contours in the left panel show the line-free K band continuum
of TNJ0121+1320. The total K band magnitude extracted from a
3\arcsec$\times$2\arcsec\ aperture in RA and dec respectively, covering
components A and B is K$=19.1\pm 0.3$ mag, consistent with the K=18.8 mag
measured by \citet{debreuck02}. After correcting for contamination with the
luminous [OIII]\lam4959,5007 emission lines from component $A$, we find
K$=19.8\pm 0.4$ mag. Note that the spatial extent of the continuum and line
emission are very similar.

Estimating the contribution of the continuum in component $B$ is more
difficult, because our spectra are not deep enough to detect this
component in the continuum.
Comparing the measured
flux of H$\beta$ and [OIII]$\lambda$4959,5007 from component $B$ in our
spectrum, F$_{[OIII],H\beta}= 3.7\times 10^{-17}$ erg s$^{-1}$ cm$^{-2}$, with
the 
K-band flux density of component $B$ measured in a 0.6\arcsec\ aperture from
the K-band image of \citet{debreuck04}, F$_{K,BB}= 4.7\times 10^{-20}$ erg
s$^{-1}$ cm$^{-2}$ 
\AA$^{-1}$ (or 22.3 mag) we find that the line emission contributes at most
20\% to the total K-band flux measured from the image. Note that we scaled the
K-band broad band data to the magnitude measured with SINFONI to ensure that
both datasets have the same calibration. Without this correction, the emission
line flux would contribute $< 15$\% to the K-band magnitude of component $B$.

Contours in the right panel of Fig.~\ref{fig:0121images} indicate the 8.4 GHz
non-thermal radio emission of TNJ0121+1320. The radio source is unresolved and 
is centered on component $A$ (\S\ref{ssec:alignment}).

\subsection{Integrated spectra of components A and B, and gas kinematics}
\begin{figure} 
\epsfig{figure=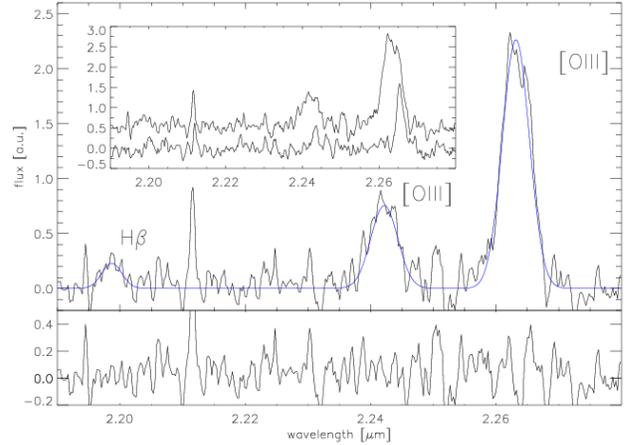,width=0.48\textwidth}
\caption{Upper panel: Integrated K-band spectrum of component A of
TNJ0121+1320. The blue line indicates the Gaussian fit to the line profile,
fit residuals are shown in the lower panel. The inset shows the integrated
spectrum of component B and the Gaussian fit. The spectrum of component A is
shown for comparison, and shifted along the ordinate by an arbitrary
amount. All spectra are extracted from 1\arcsec square apertures.} 
\label{fig:0121intspec}
\end{figure}

We extracted integrated spectra of each component from
1.0\arcsec$\times$1.0\arcsec\ apertures.
The large panel of Fig.~\ref{fig:0121intspec} shows the spectrum of the major
component $A$. [OIII]\lam4959,5007 line emission is detected
and is fitted with a Gaussian at a redshift z$=3.52001\pm
0.0007$ and a line width FWHM$=724\pm 44$ km s$^{-1}$, corrected for
instrumental resolution. We also detect \hb\ at a redshift z$=3.52293\pm
0.0008$ and a width of FWHM$=394\pm 55$ km s$^{-1}$. 
[OIII]$\lambda$5007 line emission in component B has a velocity offset
relative to component $A$ of $297\pm 7$ km s$^{-1}$, and a width of
FWHM=$274\pm 21$. 

\begin{figure} 
\epsfig{figure=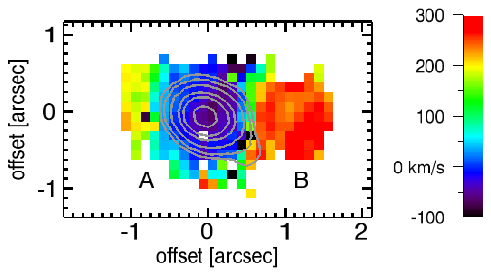,width=0.23\textwidth}
\epsfig{figure=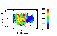,width=0.23\textwidth}
\caption{Velocity map (left) and map of the FHWMs (right) of TNJ0121+1320. 
The contours indicate the 8.4 GHz radio morphology. Colorbars show relative
velocities in km s$^{-1}$ and FWHMs in km s$^{-1}$, respectively. North is up,
east to the left in both images.}
\label{fig:0121kinematics}
\end{figure}

We used [OIII]$\lambda$5007 extracted from $0.375\arcsec \times0.375\arcsec$
apertures (3 pixels $\times$3 pixels) to trace the velocities and dispersion
of the emission line gas. The maps are shown in
Fig.~\ref{fig:0121kinematics}. As seen in the integrated spectrum, components
A and B have relative velocites of $\sim 200$ km s$^{-1}$, dispersions in
component B are significantly smaller than in component $A$. However, each
individual component is not well resolved spatially. Faint, narrow line
emission is marginally detected towards the North-West of component $A$.

\subsection{Morphology and kinematics of the ionized and molecular gas} 
\label{ssec:t0121.coopt}
\citet{debreuck03} observed CO(4--3) line emission in TNJ0121+1320, and
measured the redshift and FWHM of the molecular emission, z$_{CO} =3.520$, and
FWHM$_{CO} = 700$ km s$^{-1}$, respectively. Within the uncertainties of our
measurements, and in particular the low signal-to-noise ratio of the CO data,
the 
[OIII]$\lambda$5007 redshift and line width are identical, z$_{[OIII]} =
3.52001 \pm 0.0007$, and FWHM$_{[OIII]} = 724 \pm 44$. Similarly, both
CO(4--3) and [OIII]$\lambda$5007 line emission appear to also originate from
similarly compact regions of the galaxy: The CO emission line region is not
spatially resolved, so we use the FWHM of the beam as an upper limit to the
size, FWHM$_{CO} \le 0.54$\arcsec, which corresponds to $\le 4$ kpc. Since the
emission line region of TNJ0121+1320 A is not spatially resolved, the size of
the [OIII]$\lambda$5007 is dominated by the size of the seeing disk. This
implies an upper limit to the size of the line emitting region of of
0.5\arcsec, or $\sim 4$ kpc.

\subsection{Dynamical and molecular mass}
\label{ssec:tnj0121mass}
We will now use the smaller component {\it B} of TNJ0121+1320 to estimate a
dynamical mass for a high-redshift radio galaxy.  This component has all the
properties expected from a companion galaxy: The K-band
emission from this source is dominated by the continuum
(\S\ref{ssec:0121morph}), and the emission lines are spectrally resolved
with a velocity, $\sigma_{B}^{0121} =72\pm11$ km s$^{-1}$.  The large
difference in velocity dispersion and the velocity offset between component
$A$ and $B$ of $297\pm 7$ km s$^{-1}$ make it unlikely that they are part of
the same galaxy, although the kinematics suggest that they are physically
related, and gravitationally bound. \citep[We note that
TNJ0121+1320 $B$ appears overall very similar to the smaller component $J2$ of
the $z=2.6$ submillimeter 
galaxy SMMJ14011+0252,][]{nesvadba07b}. If the dispersion is dominated by the 
large-scale gravitational potential of TNJ0121+1320 $B$, then we can use the
dispersion to estimate a dynamical mass of component $B$ from equation
\ref{eqn:dispmass}, $M^{0121}_B \sim 7\times 10^9 M_{\odot}$.

The relative velocity between components $A$ and $B$ allows a rough dynamical
mass estimate for component $A$. The position of the radio source, the peak of
the CO line emission, and the brighter continuum emission all favor
TNJ0121+1320 A to be the more massive component.

We therefore assume that $B$ is orbiting $A$ with a circular velocity
$v_{c,A,B}=297\pm 10$ km s$^{-1}$ similar to the observed velocity offset
$v_{obs}$, and estimate the mass of component $A$, $M_{A}^{0121}$, setting
\begin{eqnarray} 
\begin{small}
M^{0121}_{A} = \frac{v_{obs}^2\ D_{A,B}}{G} = 2.3\times 10^9\
  \biggl(\frac{V}{100 {\rm km 
  s}^{-1}}\biggr) ^2 \frac{R}{1 \rm kpc} M_{\odot}
\end{small}
\end{eqnarray} where $D_{A,B} = 10$ kpc is the projected distance between the
two components. Not correcting for inclination, we estimate a dynamical mass
M$^{0121}_{A}= 2\times 10^{11}$ M$_{\odot}$.  Strictly speaking, this
is only a lower limit; due to the unknown inclination and separation, the
intrinsic mass might be factors of a few higher. For example, for the average
of a statistical sample, we expect the intrinsic masses to be larger by factors
$\sim 2-3$.

TNJ0121+1320 is also part of the sample of \citet{seymour06} with photometric
mass estimates (see \S\ref{ssec:tnj0205mass} for a discussion of their stellar
mass estimate for TNJ0205+2242); they find $M_{stellar}^{0121} \sim 1.0\times
10^{11}$ $M_{\odot}$ (in total for components $A$ and $B$), which is close to
our 
dynamical mass estimate, although the good agreement may be to some degree
fortuitous, given the large uncertainties in either estimate. Given that
component B is much fainter and appears to 
have a significantly smaller dynamical mass than component $A$, we assume that
the stellar mass budget is dominated by component $A$. Comparing with the
average stellar mass of the \citeauthor{seymour06} sample, $M_{stellar} \sim
3\times 10^{11}$ M$_{\odot}$, TNJ0121+1320 appears to have a typical mass for
HzRGs, unlike TNJ0205+2242, which appears to be somewhat less massive.

The CO(4--3) measurement of \citet{debreuck03}, $S_{43} v = 1.2$
Jy km s$^{-1}$, allows us to estimate a molecular gas mass. Following
\citet{downes98}, we adopt $M_{CO} = 0.8 {\cal L}_{CO}^{\prime}$,
where ${\cal L}^{\prime}$ is the CO luminosity in units of K km
s$^{-1}$ pc$^2$, and find a molecular gas mass of M$_{mol}^{0121}
\sim 4\times 10^{10}$ M$_{\odot}$. We note that this mass estimate is
similar to CO mass estimates of galaxies with extended radio sources
\citep[e.g.][]{papadopoulos00}.  TNJ0121+1320 is gas rich, which a
molecular mass fraction of $\sim$20\%.

\section{A unified view of AGN feedback in extended and compact high-z radio
  galaxies} 

\subsection{Constraints from optical emission line regions}

Integral-field spectroscopy of the rest-frame optical emission line gas of
z$\ge 2$ radio galaxies with extended radio sources (D$\ge$20 kpc) shows
strong evidence for enormous outflows of ionized gas
\citep{nesvadba06,nesvadba07}. These outflows have total kinetic energies of
$\sim 10^{60}$ ergs that
appear sufficient to unbind a significant fraction of the ISM of a massive,
gas-rich galaxy within the $\sim 10^7$ yrs lifetime of a radio jet.
Timescale, energy, and geometrical arguments indicate that the outflows are
most likely driven by the mechanical energy input of the radio jet, through a
mechanism by which the jet couples efficiently to the ISM of the host galaxy.
A good candidate mechanism is inflation of an overpressurized cocoon of hot
gas energized by and forming around the radio jet, that entrains and
accelerates the ambient gas as it expands
\citep{begelman89,pedlar85}. \citet{capetti99} argue that the morphology and
kinematics in the narrow-line region in the low-redshift Seyfert galaxy MRK~3
on scales of a few 100 pc are most likely powered by a inflating cocoon. Such
a mechanism would explain the high coupling efficiency that appears necessary
to explain radio emission on small scales \citep[see also. e.g.,][]{axon00,
odea02} and on larger scales \citep{nesvadba06,nesvadba07}, as well as the
ionization structure of radio galaxies at z$\sim 1$
\citep[][]{inskip02a,best00}.

This basic, and certainly over-simplified, scenario finds support from the
galaxies with compact radio sources we discuss here. Most obviously, and
similar to previous longslit studies of compact HzRGs
\citep[][]{jarvis01,inskip02a,inskip02b}, we do not 
find bright, extended emission line regions in the compact radio galaxies,
that would extend beyond the size of the radio lobes. For any mechanism where
the gas is being accelerated by the radio jets, this is certainly reassuring,
since the {\it absence} of large-scale outflows is what is expected if the
radio jet is the dominant driver of the giant outflows seen in the extended
sources, but we may see the compact sources before the jets have driven out
sufficiently large gas masses. In the past years, evidence has accumulated
that compact radio sources are young relative to extended sources
\citep[][based on their spectral properties and direct observations of
 motion within the jets]{owsianik98,murgia99,best00}, but this basic argument
does not explicitly rely on this.

\subsection{Evidence for quiescent Ly$\alpha$ halos beyond the radio lobes}
\label{ssec:lya}
\begin{figure}
\epsfig{figure=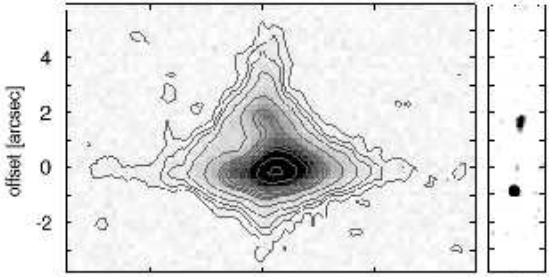,width=0.45\textwidth} 
\caption{{\it} Left: Spatially-resolved, two-dimensional longslit spectrum of
the \lya\ emission line of TNJ0205+2242. The dispersion axis is along the 
abscissa. {\it Right}: 8.4 GHz VLA map of TNJ0205+2242. Note that the size of
the broad \lya\ emitting region corresponds approximately to the size of the
radio jets, while the gas beyond the radio jets becomes more quiescent, with
FHWM$=531 $ km s$^{-1}$ and a nearly Lorentzian profile.}
\label{fig:tnj0205Lya}
\end{figure}

In Fig.~\ref{fig:tnj0205Lya} we compare the radio map of TNJ0205+2242 with the
two-dimensional rest-frame \lya\ spectrum. 
As discussed by \citet{debreuck01}, Ly$\alpha$ emission extends
over $\sim 8$\arcsec. The size of the inner, perturbed, region (with
FWHM$=$1550 km s$^{-1}$ compared to FWHM$=$531 km s$^{-1}$ in the outer
region) corresponds to the size of the radio jets, which in turn mark the
spatial extent of the bright [OIII]$\lambda5007$ emission line regions in
Fig.~\ref{fig:0205kinematics}. This good correspondence between kinematics,
and emission line surface brightness with the extent of the radio jet provides 
robust evidence that the gaseous outflows of high-redshift radio galaxies are
indeed driven by the radio jet. Note that [OIII] does not suffer from
scattering and the optical depth effects, that make \lya\ a difficult tracer 
of the kinematics. 

The \lya\ spectrum of TNJ0205+2242 suggests that compact, steep-spectrum radio
galaxies are surrounded by very similar halos as extended radio
sources. \citet{villar02,villar03,villar07} identify $\le$100-kpc sized,
kinematically quiescent \lya\ halos surronding z$\sim 2-3$ radio galaxies with
extended radio sources, which have FWHM$\sim$ 500 km s$^{-1}$ and are outside
the radius of influence of the radio jets. 

Finding a quiescent \lya\ halo around a compact HzRGs may indicate that these
galaxies reside within similarly gas rich, dense environments as extended
HzRGs, as traced by the kinematics and size of their turbulent and quiescent
\lya\ halos, adding an environmental argument to the interpretation that they
are their younger analogs.

\section{AGN feedback energy in compact radio galaxies}
\label{sec:compagnfb}
Having direct measurements of the dynamical and molecular gas
mass (\S\ref{ssec:tnj0121mass}) and jet power of TNJ0121+1320
(\S\ref{sec:introduction}), we can investigate explicitly, whether
the radio jet has the potential to unbind a significant fraction of
the ISM from the dark matter halo of the host galaxy. 

As in \citet{nesvadba06}, we use the scalings of
\citet{bullock01} to estimate that the gas will have to be accelerated to
an escape velocity of $\sim 700$ km s$^{-1}$, if it is to be unbound. This
corresponds to a total kinetic energy injection of $E \sim 9.9\times 10^{58}
M_{\odot,10}\ v_{esc,1000}$ erg, 
that must be supplied by the radio jet. $M_{\odot,10}$ is the total
mass of the outflow in units of $10^{10} M_{\odot}$, the escape velocity,
$v_{esc,1000}$, is in units of 1000 km s$^{-1}$. 

For M$_{mol}^{0121} \sim 4\times 10^{10}$ M$_{\odot}$, this corresponds to a
total energy injection of $\sim 4\times 10^{59}$ erg. Can this energy be
provided by the radio jet? Similar to \citet{nesvadba06}, we calculate the jet
kinetic energy from the jet luminosity in the frequency range of $0.1-1$ GHz
in the rest-frame, using the radio fluxes measured by \citet{debreuck01b}, and
find ${\cal L}_{jet} = 2.9\times 10^{44}$ erg s$^{-1}$. Following
\citet{deYoung93,bicknell97}, we estimate that the jet luminosity corresponds
to about 0.1-10\% of the jet kinetic energy. We adopt the second value for a
conservative estimate of the jet input energy, $\dot{E}_{kin,jet} \sim 3\times
10^{45}$ erg s$^{-1}$, or ${E}_{jet,total} \sim 9\times 10^{59}$ erg in a
$\sim 10^7$ yrs total jet lifetime. Thus, at face value, the jet may remove
the entire molecular gas reservoir from the dark matter halo, if the jet
kinetic energy is translated into kinetic energy of the gas with $\ge 40$\%
efficiency.  We estimated a lower bound to the jet kinetic energy, and
therefore may have
overestimated the actual coupling efficiency by factors of a few.
In the extended sources, we find efficiencies of on-order ${\cal O}(10\%)$,
with considerable scatter. 

Unfortunately, we do not have a direct measurement of the molecular gas
mass for TNJ0205+2242. However, given the overall similarity of the two
sources, in particular the similar radio luminosity ($2.6\times 10^{44}$
erg s$^{-1}$), suggests that any radio-jet related feedback process
will be overall very similar. Moreover, TNJ0205+2242 appears to have a
smaller mass than TNJ0121+1320 by factors of a few, so that it may have
a somewhat more shallow gravitational potential and a smaller gas mass
(\S\S\ref{ssec:tnj0205mass},\ref{ssec:tnj0121mass}).

\section{What drives the gas dynamics in compact HzRGs?}
If the radio jets in compact HzRGs are powerful enough to induce similar
outflows to those in extended radio sources, does this imply that their gas
dynamics are also dominated by the radio jet? Basic causality arguments
suggest that this is rather unlikely. Dynamical timescales of the radio jets
in compact sources, $\le 10^{5-6}$ yrs, are less than the sound crossing time
of the host (${\cal O}( 10^8 {\rm yrs}$) for a sound speed of ${\cal O}(10$ km
s$^{-1}$)).

Comparing with the mass outflow rates in extended HzRGs, we can give an
order-of-magnitude estimate of the entrainment rate, and find a similar
answer. Since the entrainment rate will depend crucially on the gas content
in the host galaxy, which seems to evolve strongly with redshift
(\S\ref{sec:introduction}), we cannot rely on entrainment rates estimated from
detailed jet modelling in compact radio sources at low redshift
\citep[e.g.][]{laing02}. Assuming an overall constant entrainment rate within
a factor $\sim 10$ during the jet lifetime, and using the observed $dM/dt
\sim$ few $\times 100\ M_{\odot}$ yr$^{-1}$, corresponding to the estimates of
\citet{nesvadba06, nesvadba07}, a simple scaling can approximate the total
mass of entrained material,

\begin{equation}
M_{entr} = f_{c}\ f_{\rho_r}\ \frac{\dot{M}_{obs}}{M_{\odot} yr^{-1}}\
\frac{\tau}{10^6 yrs}\  [10^8 M_{\odot}]
\end{equation}

with the outflow rate $\dot{M}_{obs}$ observed in extended radio galaxies
\citep{nesvadba06, nesvadba07}, and age of the radio source, $d\tau$. $f_{c}$
and $f_{\rho(r)}$ are correction factors accounting for the radial density
profile of the host galaxy and variations in the jet advance speed,
respectively. \citet{readhead96} argue that the jet advance speed is probably
not a strong function of radius, so that $f_{c} \sim 1$. For approximately
uniform density, $\rho(r) \sim \rho_0$, we find that after $10^6$ yrs, only
few $\times 10^8 M_{\odot}$ will have been entrained by the radio jet, less
than a percent of the $4\times 10^{10}$ M$_{\odot}$ in molecular mass observed
in TNJ0121+1320. The mass outflow rate in extended HzRGs was estimated from
the mass of ionized gas only, and certainly underestimates the intrinsic mass
outflow rate by factors of a few. Similarly, a non-uniform density profile
will imply a larger entrainment rate in young radio sources. Therefore, we
likely underestimate the true entrainment rate by factors of a few. However,
even for a $\sim 10 \times$ larger entrained gas mass, no more than $\sim
10$\% of the molecular gas mass could have been entrained in a compact
HzRG. This is a strong contrast to extended HzRGs, which have entrained gas
masses similar if not greater than their molecular gas content
\citep{nesvadba07}, and may be a simple consequence of the younger ages of
compact sources.  

There is an interesting trend if we compare the two sources, although
statements based on only two galaxies can hardly be conclusive. Using the size
of the radio jet to estimate the age of each source, we find that TNJ0205+2242
may be factors of a few larger than the more compact TNJ0121+1320
\citep[t$^{0205}_{jet}\le 4\times 10^{6}$ yrs 
yrs compared to t$^{0121}_{jet}\le 7\times 10^5$ yrs, for TNJ0121+1320 and
TNJ0205+2242, respectivley, and for a jet advance speed of v$\ge 0.01c$,
following][]{wellman97}. A similar
trend is seen in the emission line properties of the two sources: 
TNJ0205+2242 has a biconal component that appears related to the radio jet, as
suggested by size and spatial alignment, although it has a smaller velocity
gradient , $\sim 200$ km s$^{-1}$, than extended radio
galaxies \citep[$\Delta v \ge 1000$ km s$^{-1}$][]{nesvadba07}, and 
contains only a few percent of the total ionized gas, if the 
[OIII]$\lambda$5007/H$\beta$ line ratios are roughly constant. This galaxy
also has an unresolved, broad [OIII]$\lambda$5007 emission line, which
is blueshifted by $\sim 200$ km s$^{-1}$ relative to the narrow [OIII]
component, and has a line width FWHM$\sim 1200$ km s$^{-1}$ similar to
extended radio sources.  This may indicate a similarly
turbulent, but less evolved, outflow as seen in the extended
HzRGs. TNJ0121+1320 has more narrow lines (FWHM$\sim 700$ km s$^{-1}$, 
which is not unplausibly high for a galaxy with a mass of a few $\times
10^{11}$ M$_{\odot}$), and no biconal component.
Note that the turbulent line broadening will be mostly isotropic, hence
we do not expect differences in viewing angle to have a large impact.

Overall, the properties of each of these two sources are in broad agreement
with what would be expected if the gas dynamics of the host galaxies were more
and more dominated by the energy injection of the radio jet as the radio
source ages and expands: TNJ0205+2242, which has a more extended, and
presumably more evolved radio source, resembles galaxies with well extended
radio sources more than TNJ0121+1320, whose radio source is very compact, and
presumably younger. 

\section{Summary}
We presented a study of the rest-frame optical emission line gas of
two powerful radio galaxies at z=3.5, TNJ0205+2242 and TNJ0121+1320,
using integral-field spectroscopy, combined with radio, rest-frame UV,
and CO emission line data. Both galaxies have compact radio emission, with
radii $\le$ 10 kpc. They have bright [OIII]$\lambda$4959,5007 and H$\beta$
line emission, and faint, compact continuum emission. The excessive
emission line equivalent widths imply that line contamination is a serious
complication for studying the continuum morphologies of high-redshift
radio galaxies at redshifts where strong line emission falls within
the bandpass of broad band filters, and adds additional uncertainties in
obtaining SEDs of high-redshift radio galaxies. Both galaxies have a broad
[OIII]$\lambda$5007,4959 emission line component, with FWHMs $\sim 700$
km s$^{-1}$ and $\sim 1200$ km s$^{-1}$ for TNJ0121+1320 and TNJ0205+2242
respectively. TNJ0205+2242 has a narrow component superimposed, which
is redshifted relative to the broad component by $\sim 200$ km s$^{-1}$.

A secondary component of discrete emission in TNJ0121+1320
offers a unique possibility to estimate the dynamical mass of a
high-redshift radio galaxy in the same way as often applied to other
galaxy populations. TNJ0121+1320 $B$, at a distance of $\sim 10$ kpc
from the massive radio galaxy, and at a velocity offset of $\sim 300$ km
s$^{-1}$, has all the properties of being a M$ \sim 10^{9-10} M_{\odot}$
galaxy that is physically connected to the more massive component, and may be
in 
the process of merging.  From the velocity offset and projected distance,
we estimate that the major component of TNJ0121+1320 has a dynamical
mass of $\sim 2\times 10^{11}$ M$_{\odot}$ within a radius of $\sim 10$
kpc. This agrees well with photometric mass estimates of TNJ0121+1320
\citep{seymour06}, and suggests that TNJ0121+1320 has a typical mass for
high-redshift radio galaxies. For TNJ0205+2242, we estimate a somewhat
lower mass  of a few  $\times 10^{10}$ M$_{\odot}$ from the dispersion of
the narrow [OIII]$\lambda$5007 component, using the empirical relationship
between FWHM of the narrow-line region and the stellar velocity dispersion
of \citet{nelson96}. This again agrees well with the photometric mass
estimate of \citet{seymour06}, who find 4 $\times 10^{10}$ M$_{\odot}$.

Emission line morphologies in both sources are compact, and do not extend
beyond the most extended radio emission. This is a necessary condition
for arguing that radio jets may provide a sufficiently efficient coupling
mechanism between the AGN and the ISM with the potential of removing
significant gas masses from the host galaxy. In particular, this is fully
compatible with the conclusion that the large-scale energetic outflows
observed in extended radio galaxies appear to be powered predominantly
by the radio jet \citep{nesvadba06,nesvadba07}. Similarly, compact radio
galaxies have jets that appear powerful enough to unbind significant
gas fractions from a massive, gas-rich galaxy, and specifically, to
unbind the 4$\times 10^{10} M_{\odot}$ in molecular mass estimated
from CO(4--3) measurements in TNJ0121+1320. Adopting the ``paradigm of
youth'', that the compactness of a radio source is a sign of young age,
and estimating entrainment rates from the large-scale outflows we observe in
galaxies with extended radio sources, we find that compact radio sources
have not yet had the time to entrain significant gas masses. Timescale
and causality arguments, as well as the individual properties of the
likely more evolved TNJ0205+2242 relative to the likely somewhat younger
TNJ0121+1320, are broadly consistent with this picture.

\acknowledgements
The authors wish to thank the anonymous referee for good advice on how to
improve the paper. We are grateful to the ESO OPC for their allocation of
observing time and the staff at Paranal for their help and support in making
these observations.  NPHN wishes to acknowledge financial support from the
European Commission through a Marie Curie Postdoctoral Fellowship and MDL
wishes to thank the Centre National de la Recherche Scientifique for its
continuing support of his research. Part of this work was performed under the
auspices of the U.S. department of Energy, National Nuclear Security
Administration by the University of California, Lawrence Livermore National
Laboratory under contract No. W-7405-Eng-48. WvB acknowledges support for
radio and infrared galaxy studies with the Spitzer Space Telescope at UC
Merced, including the work reported here, via NASA grants SST GO-1264353,
GO-1265551, GO-1279182 and GO-1281587.

\clearpage
\begin{table}
\caption{Emission Lines in TNJ0205+2242}
\label{tab:exposure}
\centering
\begin{tabular}{lccccc}
\hline \hline
Line & $\lambda_0$ & $\lambda_{obs}$ & z & FWHM & flux \\
(1)  & (2)         & (3)            & (4)& (5) & (6) \\
\hline
$[$OIII$]$ & 5007 & 2.25754$\pm$0.00045 & 3.50867$\pm$0.00070 & 143 $\pm$10  & 2.7 $\pm$ 0.19 \\
$[$OIII$]$ & 5007 & 2.25608$\pm$0.00045 & 3.50574$\pm$0.00070 & 1171$\pm$72  & 23.5$\pm$ 1.4 \\
H$\beta$   & 4861 & 2.19179$\pm$0.00054 & 3.50882$\pm$0.00086 & 1699$\pm$145 & 3.8 $\pm$ 0.37 \\ 
\hline
\end{tabular}
\caption{ Column (1) -- line. Column (2) -- Rest-frame
  wavelength. Column (3) -- Observed wavelength. Column (4) --
  Redshift. Column (5) -- FWHM, corrected for instrumental resolution. Column
  (6) -- Line flux in units of 10$^{-17}$ erg s$^{-1}$ cm$^{-2}$.}
\end{table}

\begin{table}
\caption{Emission Lines in TNJ0121+1320}
\label{tab:exposure}
\centering
\begin{tabular}{lcccccc}
\hline \hline
Component & Line & $\lambda_0$ & $\lambda_{obs}$ & z & FWHM & flux \\
(1)  & (2)         & (3)            & (4)& (5) & (6) & (7) \\
\hline
A &$[$OIII$]$  & 5007 & 2.26322 $\pm$ 0.00045 & 3.52001$\pm$ 0.0007 & 724$\pm$ 44 & 13.2\\
A & H$\beta$   & 4861 & 2.19864 $\pm$ 0.00047 & 3.52293$\pm$ 0.0008 & 394$\pm$ 55 & 1.0 \\ 
B &$[$OIII$]$  & 5007 & 2.26541 $\pm$ 0.00032 & 3.52449$\pm$ 0.0007 & 274$\pm$ 21 & 2.4 \\
B & H$\beta$   & 4861 & 2.20054 $\pm$ 0.00034 & 3.52692$\pm$ 0.0008 & 153$\pm$ 47 & 0.3\\
\hline
\end{tabular}
\caption{ Column (1) -- component. Column (2) -- line. Column (3) -- Rest-frame
  wavelength. Column (4) -- Observed wavelength. Column (5) --
  Redshift. Column (6) -- FWHM, corrected for instrumental resolution. Column
  (7) -- Line flux in units of 10$^{-17}$ erg s$^{-1}$ cm$^{-2}$.}
\end{table}

\end{document}